\documentclass[%
 reprint,
 amsmath,amssymb,amsthm,
 nofootinbib,
 aps,
]{revtex4-1}

\usepackage{dcolumn}
\usepackage{bm}
\usepackage[mathlines]{lineno}
\usepackage{graphicx}
\usepackage{array}
\usepackage{csquotes}

\begin{document}
\preprint{APS/123-QED}

\title{Quantum Entanglement Induced by Gravitational Potential}

\author{J. Michael Yang}
\email{jianhao.yang@alumni.utoronto.ca}
\affiliation{
Qualcomm, San Diego, CA 92121, USA
}

\date{\today}

\begin{abstract}
\textbf{Abstract:} The purpose of this study is to calculate the entanglement measure for a bipartite system where the two subsystems interact via a central potential, and more importantly, to analyze the conceptual implication in the case of gravitational potential. Through numerical calculation, we confirm that the ground state of such quantum system is an entangled state. This raises a question whether a quantum ground state describing masses interacting through gravity can ever be a separable state. Entanglement seems to be an intrinsic characteristic for the ground state when mutual gravitational interaction is involved. Although the study is in the context of non-relativistic quantum mechanics, it provides hints for a quantum gravity theory in the limit of very weak field and low relative velocity. The entanglement of the ground state of two masses seems intrinsically connecting to the curvature of spacetime they create. 
\begin{description}
\item[Keywords] Quantum Entanglement, Purity, Gravitational Interaction, Geometry of Spacetime
\end{description}
\end{abstract}
\maketitle

\section{Introduction}
\label{intro}
Quantum entanglement has been a source of theoretical interest in probing the foundation of quantum mechanics. For instance, it was used in EPR thought experiment to argue that quantum mechanics is an incomplete physical theory~\cite{EPR}. Quantum entanglement is considered a more fundamental property in some of the quantum mechanics interpretations such as decoherence theory~\cite{Zurek82, Zurek03} and relational quantum mechanics~\cite{Rovelli96, Rovelli07, Yang2017, Yang2017_2}. In recent decades, it is recognized that entanglement is connecting to the gravitational dynamics in the context of holography. Holography in high energy physics refers to the duality, or more specifically, the gauge/gravity or AdS/CFT correspondence~\cite{Maldacena,Ryu,Raamsdonk}. Entanglement as an unique communication resource has been widely explored in recent communication technology~\cite{Nielsen00,Hayashi15}.

In this paper we wish to study the entanglement of two quantum systems interact through central potential, and particularly, the conceptual implication of such entanglement properties. Although the entanglement properties of such bipartite systems have been studied before in literature~\cite{Cohen, Tommasini, Marcelo}, there are important conceptual implications that are not fully recognized. Here we briefly discuss one possible implication. 

Although the idea that entanglement is related the emergence of spacetime is inspiring, the connection is indirect via the AdS/CFT correspondence. In this setting, the emergence of geometry occurs in a $d+1$ dimensional AdS with gravity, while the entanglement is between a spatial region in a $d$ dimensional boundary of AdS and the rest of the boundary without gravity. It is natural to ask whether there is direct connection between entanglement and the emergence of spacetime geometry without the need of the holographic correspondence. Furthermore, one may also wonder if the connection in the holographic context is a unique result from the string theory, or it is intrinsic to any plausible quantum gravity theory. Unfortunately, these questions cannot be easily answered as we do not have a truly unified quantum gravity theory. Practically, we can take one step back and ask a simpler question: can the traditional quantum theory provide similar hints on the connection between entanglement and gravity? To probe the answer to this question, one can investigate whether entanglement between two massive objects can be induced through interaction of classical gravity. The classical result can be seen as a constraint for the unified theory in the limit of very weak field and low relative velocity. 


There are other possible implications we can draw from the study of entanglement of such bipartite system. For instance, is the entanglement a property depending on the reference frame, or it is invariant when switching reference frames? Given the ubiquitous of Coulomb interaction and gravity interaction, is entanglement a ubiquitous quantum phenomenon?

With these motivations, we first develop a generic formulation to calculate the reduced density matrix of a bipartite system with continuous variable. This allows us to derive an entanglement measure based on purity of the reduced density matrix. The formulation is then applied to calculate the entanglement measure of the ground state of a bipartite system where the two subsystems interact through a central potential. The numerical results show that the ground state of the two masses is an entangled state, so as the first excited state. 
Although our calculation appears straightforward and confirms result from earlier studies~\cite{Tommasini}~\footnote{We would like to point out that the calculation of entanglement measure in Ref.~\cite{Tommasini} is formulated only for a particular class of states that are translationally invariant. To the contrary, the formulation to calculated the entanglement measure is generic for any bipartite pure state including those states without translational invariance.}, our primary goal is to analyze the conceptual implication. First it raises an interesting question whether a quantum ground state describing masses interacting through gravity can ever be a separable state. Entanglement seems to be an intrinsic characteristic for the ground state when mutual gravitational interaction is involved. It suggests that the entanglement manifests the curvature of spacetime if we take the general relativity perspective. It can be seen as a check point for a unified theory in the limit of very weak field and low relative velocity. However, we should point out that this paper does not yet give a definite answer on how entanglement is connected to the geometric properties of spacetime, as the question can only be answered in a unified quantum gravity theory. Instead, this paper is intended to provide hints for the pursuit of answering such question. Other implications, such as the dependency of entanglement on reference frame can also be confirmed from our calculation.


The paper is organized as following. In Section \ref{sec:entmeasusre} we derive a generic formulation of entanglement measure for a bipartite system with continuous variable. The formulation is then applied to the ground state derived from the Schr\"{o}dinger equation with gravity potential in Section \ref{sec:centralmass} to obtain explicit expression of entanglement measure. The numerical calculations shown in Section \ref{sec:montecarlo} clearly show that the ground state is an entangled state.  Section \ref{sec:Interferometers} is dedicated to discussing the similarity and difference between the results in Refs~\cite{Bose, Marletto} and the results in this paper. Section \ref{sec:discussion} explores the conceptual implications of our calculation results. Limitations and conclusive remarks are presented in Section \ref{sec:limit} and \ref{sec:conclusion}, respectively.

\section{Entanglement Measure for Continuous Variable System}
\label{sec:entmeasusre}
For a continuous variable bipartite system with subsystem 1 and 2, a pure state can be expressed as 
\begin{equation}
\label{WF1}
|\Psi_{12}\rangle=\int \psi(x,y)|x\rangle|y\rangle dxdy.
\end{equation}
Here we assume one dimensional continuous variable $x, y$ for each subsystem 1 and 2, respectively. It is straightforward to extend to three dimensional variables. Normalization requires
\begin{equation}
\label{Normalization}
\int \psi(x,y)\psi^*(x,y) dxdy = 1.
\end{equation}
Rewriting (\ref{WF1}) in the form of density matrix, we have
\begin{equation}
\label{rho1}
\begin{split}
\hat{\rho}_{12} & = |\Psi_{12}\rangle\langle_{12}\Psi| \\
& =\int\int \psi(x,y)\psi^*(x', y')|x\rangle|y\rangle\langle x'|\langle y'| dxdydx'dy'.
\end{split}
\end{equation}
This gives the density matrix element
\begin{equation}
\label{rho2}
\rho_{12}(x,y; x',y') =\psi(x,y)\psi^*(x', y').
\end{equation}
The reduced density matrix for subsystem 1 can be derived by taking partial trace,
\begin{equation}
\label{rho3}
\begin{split}
\hat{\rho}_{1} & = Tr_2(\hat{\rho}_{12}) = \int \textsubscript{2}\langle z|\Psi_{12}\rangle\langle_{12}\Psi|z\rangle_2 dz \\
& =\int \{\int \psi(x,z)\psi^*(x', z)dz \}|x\rangle\langle x'| dxdx'.
\end{split}
\end{equation}
Here, the integration over variable $z$ is on subsystem 2. From (\ref{rho3}), one obtains the reduced density matrix element for subsystem 1,
\begin{equation}
\label{rho4}
\rho_{1}(x; x') =\int \psi(x,z)\psi^*(x', z)dz.
\end{equation}
Due to the normalization property in (\ref{Normalization}), we have
\begin{equation}
\label{Trace1}
\begin{split}
Tr(\hat{\rho}_{1}) & =\int \rho_{1}(x; x)dx \\
& = \int \psi(x,z)\psi^*(x, z)dzdx  = 1,
\end{split}
\end{equation}
as expected. To quantify the entanglement between subsystem 1 and 2, we use the following definition~\cite{Calmet}
\begin{equation}
    \label{ent}
    E = 1 - Tr(\hat{\rho}_1^2).
\end{equation}
Quantity $Tr(\hat{\rho}_1^2)$ is the purity of reduced density matrix $\rho_1$, given by
\begin{equation}
\label{purity}
\begin{split}
Tr(\hat{\rho}_1^2) &=\int \rho_1^2(x,x)dx \\
&= \int \rho_1(x, x')\rho_1(x', x)dx'dx.
\end{split}
\end{equation}
Appendix A shows the justification on why $E$ can be considered as an entanglement measure. Substitute (\ref{rho4}) into (\ref{purity}) and replace variable $z$ with $y$, we have
\begin{equation}
    \label{purity2}
    \begin{split}
    &Tr(\hat{\rho}_1^2) = \\ &\int \psi(x,y)\psi^*(x', y)\psi(x',y')\psi^*(x, y') dydy'dx'dx.
    \end{split}
\end{equation}
Thus, given a continuous variable wave function of a bipartite system, $\psi(x,y)$, we can calculate the entanglement measure from (\ref{ent}) and (\ref{purity2}). Suppose that the wave function can be factorized as $\psi(x,y)=\phi(x)\varphi(y)$,
\begin{equation}
    \label{purity3}
    Tr(\hat{\rho}_1^2) = (\int |\phi(x)|^2 dx\int |\varphi(y)|^2dy)^2 = 1.
\end{equation}
Thus, $E=0$, there is no entanglement between the two subsystems. However, when $\psi(x,y) \ne \phi(x)\varphi(y)$, it is not obvious whether $E\ne 0$. A detailed calculation is needed.

We can extend (\ref{purity2}) to three-dimensional system with continuous variable. Suppose the continuous variables for subsystem 1 and 2 are $\vec{r}_1$ and $\vec{r}_2$, respectively, then
\begin{equation}
\label{purity4}
\begin{split}
    Tr(\hat{\rho}_1^2) &= \int \psi(\vec{r}_1,\vec{r}_2)\psi^*(\vec{r'}_1, \vec{r}_2)\psi(\vec{r'}_1,\vec{r'}_2) \\ 
    & \times \psi^*(\vec{r}_1, \vec{r'}_2) d\vec{r'}_2d\vec{r}_2d\vec{r'}_1d\vec{r}_1,
\end{split}
\end{equation}
where $d\vec{r}:=dxdydz$. Note that (\ref{purity4}) is a twelve dimensional integration.

\section{Entanglement of Two Masses Interacting With Classical Gravity}
\label{sec:centralmass}
Consider a three-dimensional bipartite system, and the interaction between two subsystems is described as a potential only depends on the distance between the two subsystems $r_{12} = |\vec{r}_1-\vec{r}_2|$. The stationary Schr\"{o}dinger Equation is given by
\begin{equation}
    \label{SE}
    [-\frac{\hbar^2}{2m_1}\nabla^2_1 -\frac{\hbar^2}{2m_2}\nabla^2_2 + V(r_{12})]\psi(\vec{r}_1,\vec{r}_2) = E_t\psi(\vec{r}_1,\vec{r}_2),
\end{equation}
where $\nabla^2$ is the Laplacian operator in the three-dimensional coordinator, $E_t$ is the total energy of the system, $m_1$ and $m_2$ are the masses of subsystem 1 and 2, respectively. The potential can be a Coulomb potential or a gravity potential. The formulation is constructed here in the context of traditional quantum mechanics. Nevertheless, this can be considered as a first order approximation of more general formulation. The question we want to answer is that given a solution of wave function from (\ref{SE}), whether the two subsystems are in entangled state. In particular, we are interested in the ground state.

A typical way to solve (\ref{SE}) is to introduce transformation ~\cite{Messiah}
\begin{equation}
    \label{transform}
    \begin{split}
    \vec{r}_{12} &= \vec{r}_1-\vec{r}_2, \\
    \vec{R}_{12} & = \frac{m_1}{m_1+m_2}\vec{r}_1 + \frac{m_2}{m_1+m_2}\vec{r}_2,
    \end{split}
\end{equation}
where $\vec{R}_{12}$ is the center-of-mass coordinate of the system. Omit the subscription ``12" and denote
\begin{equation} 
    \label{transform2}
    \psi(\vec{r}_1,\vec{r}_2)=\phi(\vec{R})\varphi(\vec{r}), 
\end{equation}
(\ref{SE}) is separated into two equations ,
\begin{equation}
    \label{SE2}
    -\frac{\hbar^2}{2M}\nabla^2_R \phi(\vec{R}) = E_c\phi(\vec{R}),
\end{equation}
\begin{equation}
    \label{SE3}
    [-\frac{\hbar^2}{2\mu}\nabla^2_r + V(r)]\varphi(\vec{r}) = E_r\varphi(\vec{r}),
\end{equation}
where $M=m_1+m_2$ is the total mass, $\mu=m_1m_2/(m_1+m_2)$ is the effective mass, $E_c$ is the kinetic energy of the center mass, $r=|\vec{r}_{12}|$, and $E_r=E_t-E_c$. 

Eq.(\ref{SE2}) corresponds to the Schr\"{o}dinger equation of a free particle. Suppose the center mass of the bipartite system is moving with a constant momentum $\vec{P}_c$, and the system is in a three-dimensional spatial box with length $L$, each of the dimensional variable $x,y,z\in \{-L/2, L/2\}$. The wave function $\phi(\vec{R})$ can be expressed as
\begin{equation}
    \label{WF2}
    \phi(\vec{R}) = \sqrt{\frac{1}{L^3}}e^{i\vec{P}_c\cdot\vec{R}/\hbar}.
\end{equation}
with the understanding of taking the limitation $L\to\infty$ after integration in subsequent calculation. Substitute this into (\ref{transform2}) and then into (\ref{purity4}), the purity of reduced density matrix is simplified as
\begin{equation}
\label{purity5}
\begin{split}
    Tr(\hat{\rho}_1^2) = & \lim_{L\to\infty}(\frac{1}{L})^6\int_{-L/2}^{L/2} |\int_{-L/2}^{L/2}\varphi(\vec{r}_{12}) \\
    &\times \varphi^*(\vec{r}_{12'})d\vec{r}_1|^2 d\vec{r'}_2d\vec{r}_2.
\end{split}
\end{equation}
Once the wave function for the relative movement between the two subsystems, $\varphi(\vec{r}_{12})$, is solved from (\ref{SE3}), one can compute the entanglement $E$ from (\ref{purity5}) and (\ref{ent}).

The solution of (\ref{SE3}) depends on the actual form of central potential energy $V(r_{12})$ where $r_{12}$ is the relative distance between the two subsystem. For gravitational potential energy,
\begin{equation}
    \label{Gravity}
    V(r_{12}) = -G\frac{m_1m_2}{r_{12}},
\end{equation}
where $G$ is the Newtonian constant of gravitation. Another example is the Coulomb potential energy of hydrogen atom, given by
\begin{equation}
    \label{Coulumb}
    V(r_{12}) = -\frac{e^2}{r_{12}}.
\end{equation}
(\ref{SE3}) with such potential energies can be solved analytically. In particular, we are interested the solution for the ground state. The wave function for the ground state is given by~\cite{Saxon}
\begin{equation}
    \label{GroundState}
    \varphi_g(\vec{r}_{12}) = \sqrt{\frac{1}{\pi a^3}}e^{-r_{12}/a}.
\end{equation}
Here, constant $a$ in the case of Coulomb potential energy given by (\ref{Coulumb}) is $\hbar^2/(\mu e^2) = 0.83\times 10^{-8}cm$, which is the famous \textit{Bohr Radius}. In the case of gravity potential energy given by (\ref{Gravity}), 
\begin{equation}
    \label{GravityRadius}
    a = \frac{\hbar^2}{G\mu m_1m_2} = \frac{\hbar^2}{G m_1^2m_2}(1+\frac{m_1}{m_2}),
\end{equation}
which is called the \textit{Gravitational Bohr Radius}. We will discuss the practical meaning of this constant in section \ref{sec:limit}. For the time being, just consider it as a constant in the solution for the relative wave function $\varphi(\vec{r}_{12})$.

Substituting (\ref{GroundState}) into (\ref{purity5}), after some algebra, we obtain
\begin{equation}
\label{purity6}
\begin{split}
    Tr(\hat{\rho}_1^2) = & \lim_{L\to\infty}\frac{1}{\pi^2a^6L^6}\int_{-L/2}^{L/2} d\vec{r'}_1d\vec{r}_1 d\vec{r'}_2d\vec{r}_2 \\
    &\times e^{-2(r_{12}+r_{1'2}+r_{12'}+r_{1'2'})/a}.
\end{split}
\end{equation}
Replacing the position variable $\vec{r}$ with non-dimensional variable $\vec{\gamma} = 2\vec{r}/L$, for $\vec{r'}_1,\vec{r}_1, \vec{r'}_2, \vec{r}_2$, and denoting $\alpha=L/a$, we rewrite (\ref{purity6}) as
\begin{equation}
\label{purity7}
\begin{split}
    Tr(\hat{\rho}_1^2) = & \lim_{\alpha\to\infty}\frac{1}{\pi^2}(\frac{\alpha}{4})^6\int_{-1}^{1} d\vec{\gamma'}_1d\vec{\gamma}_1 d\vec{\gamma'}_2d\vec{\gamma}_2 \\
    &\times e^{-\alpha(\gamma_{12}+\gamma_{1'2}+\gamma_{12'}+\gamma_{1'2'})}.
\end{split}
\end{equation}
Explicitly written in Cartesian coordinate, variable $\gamma_{12}=2\sqrt{(x_1 - x_2)^2+(y_1 - y_2)^2+(z_1 - z_2)^2}/L$. Similar expressions can be written down for $\gamma_{1'2}$, $\gamma_{12'}$, and $\gamma_{1'2'}$. $d\vec{\gamma}_1=(2/L)^3 dx_1dy_1dz_1$, and similar expressions for $d\vec{\gamma'}_1$, $d\vec{\gamma'}_2$, and $d\vec{\gamma}_2$.
We can also compute the entanglement measure for an excited state. The relative wave function for the first spherical symmetry excited state is given by~\cite{Saxon}
\begin{equation}
    \label{excitedState}
    \varphi_e(\vec{r}_{12}) = \sqrt{\frac{1}{8\pi a^3}}(1-\frac{r_{12}}{2a})e^{-r_{12}/2a}.
\end{equation}
Substituting this into (\ref{purity5}), using the same notations as for (\ref{purity7}), and denoting $\beta=L/(4a)=\alpha/4$, we obtain
\begin{equation}
\label{purity8}
\begin{split}
    Tr(\hat{\rho}_1^2)_e = & \lim_{\beta\to\infty}\frac{1}{\pi^2}(\frac{\beta}{2})^6\int_{-1}^{1} d\vec{\gamma'}_1d\vec{\gamma}_1 d\vec{\gamma'}_2d\vec{\gamma}_2 \\
    &\times (1-\beta\gamma_{12})(1-\beta\gamma_{1'2})(1-\beta\gamma_{12'})\\
    &\times (1-\beta\gamma_{1'2'})e^{-\beta(\gamma_{12}+\gamma_{1'2}+\gamma_{12'}+\gamma_{1'2'})}.
\end{split}
\end{equation}
As seen, both (\ref{purity7}) and (\ref{purity8}) contain twelve dimensional integrals. There are no analytic results. Numerical calculation is needed.

\section{Monte Carlo Integration}
\label{sec:montecarlo}
In this section, Monte Carlo integration method is utilized to estimate the twelve-dimensional integration in (\ref{purity7}). We use the MISER algorithm~\cite{MISER} of GNU Scientific Library version 2.5. To validate the accuracy of the algorithm, we first calculate $Tr(\rho_1)$. It is expected to have $Tr(\rho_1)=1$ due to the normalization requirement. Followed similar derivation steps in previous section, $Tr(\rho_1)$ is expressed as
\begin{equation}
    \label{Normalization2}
    Tr(\hat{\rho}_1) =  \lim_{\alpha\to\infty}\frac{1}{\pi}(\frac{\alpha}{4})^3\int_{-1}^{1} e^{-\alpha(\gamma_{12})} d\vec{\gamma}_1 d\vec{\gamma}_2.
\end{equation}
This is a six-dimensional integration. Table \ref{tab:1} shows the calculation results with different value of $\alpha$ using the Monte Carlo integration algorithm. The MISER algorithm is set to recursively calculate the integration till the statistical error is less than one percent. Double precision variables are used in the calculation. The number of Monte Carlo calls $N_{MC}$ increases significantly when $\alpha$ increases. Our calculation ends at $\alpha=L/a = 200$. From the results in Table \ref{tab:1}, it is reasonable to extrapolate that $Tr(\rho_1)\to 1$ when $\alpha\to\infty$. This confirms the MISER algorithm is fairly accurate. In Appendix B, we have also calculated $Tr(\rho^2_1)$ analytically for a harmonic oscillator, and verified that the numerical calculation using the MISER algorithm is consistent with the analytic result. This further confirms the reliability of the Monte Carlo method for a twelve-dimension integration.
\begin{table}[h!]
\caption{Trace of Reduced Density Matrix $\rho_1$}
\label{tab:1}       
\begin{tabular}{m{1cm}|m{1.5cm}|m{1.5cm}|m{1.5cm}}
\hline
$L/a$ & $Tr(\hat{\rho}_1)$ & Error & $N_{MC}$ (million) \\
\hline
10 & 0.786360 & 0.004386 & 1\\
20 & 0.876379 & 0.008357 & 2\\
40 & 0.955348 & 0.006011 & 16\\
100 & 0.981460 & 0.006937 & 128\\
150 & 0.984406 & 0.005142 & 512\\
200 & 0.994176 & 0.007497 & 1,024\\
\hline
\end{tabular}
\end{table}

We now proceed to calculate the purity of reduced density $\rho_1$ derived from the ground state, given in (\ref{purity7}). The results are shown in Table \ref{tab:2}.
\begin{table}[h!]
\caption{Purity of $\rho_1$ at Ground State}
\label{tab:2}       
\renewcommand{\arraystretch}{1.2}
\begin{tabular}{m{1cm}|m{2cm}|m{2cm}|m{1.5cm}}
\hline
$L/a$ & $Tr(\hat{\rho}_1^2)$ & Error & $N_{MC}$ (million) \\
\hline
10 & $1.03\times10^{-4}$ & $3.89\times 10^{-6}$ & 256\\
20 & $1.60\times 10^{-5}$ & $2.53\times 10^{-6}$ & 1,024\\
40 & $5.46\times 10^{-7}$ & $9.34\times 10^{-8}$ & 2,048\\
60 & $1.55\times 10^{-8}$ & $1.34\times 10^{-9}$ & 2,048\\
100 & $1.53\times 10^{-9}$ & $2.98\times 10^{-10}$ & 2,048\\
\hline
\end{tabular}
\renewcommand{\arraystretch}{1}
\end{table}
Note that the calculation becomes very expensive when $\alpha$ increases, as the number of Monte Carlo calls increases to billions. To reduce the computation cost, the calculation is terminated whenever the error is $<$20\% of the value of $Tr(\rho_1^2)$. From the results in Table \ref{tab:2}, $Tr(\rho_1^2)\to 0$ rapidly when $\alpha$ increases. It is reasonably to extrapolate that $E=1-Tr(\rho_1^2)=1$ when $\alpha \to \infty$. This confirms that the ground state is an entangled state. The two masses are entangled due to the gravity interaction.

The Monte Carlo integration results for (\ref{purity8}) are shown in Table \ref{tab:3}. The numerical results show that $Tr(\rho_1^2)\to 0$ asymptotically when $\alpha$ increases. Compared to Table \ref{tab:2}, the purity approaching zero slower when the bipartite system is in the excite state. Nevertheless, it still shows that the two subsystems are in an entangled state.
\begin{table}[h!]
\caption{Purity of $\rho_1$ at Ground State}
\label{tab:3}       
\renewcommand{\arraystretch}{1.2}
\begin{tabular}{m{1cm}|m{2cm}|m{2cm}|m{1.5cm}}
\hline
$L/a$ & $Tr(\hat{\rho}_1^2)$ & Error & $N_{MC}$ (million) \\
\hline
10 & $4.94\times10^{-2}$ & $1.64\times 10^{-4}$ & 1\\
40 & $2.13\times 10^{-2}$ & $1.40\times 10^{-4}$ & 128\\
100 & $2.07\times 10^{-3}$ & $1.15\times 10^{-4}$ & 2,048\\
200 & $6.97\times 10^{-5}$ & $5.94\times 10^{-6}$ & 2,048\\
400 & $5.86\times 10^{-6}$ & $1.51\times 10^{-6}$ & 8,192\\
\hline
\end{tabular}
\renewcommand{\arraystretch}{1}
\end{table}

\section{Practical Limitations}
\label{sec:limit}
Although we show that two masses interacting through classical gravity potential field are entangled when they are in the ground state, there are practical limitations to detect such entanglement.

The limitation can be examined from the only numerical parameter $\alpha=L/a$ in (\ref{purity7}) which depends on physical parameter $a$. In the case of hydrogen atom, $a=0.83\times 10^{-8}cm$ is the Bohr Radius. The electron and the hydrogen nuclear are entangled when they are in the ground state. As correctly pointed out in Ref.~\cite{Marcelo}, due to the large asymmetry of mass between electron and proton, the entanglement is very small. On the other hand, for a positronium system that comprised a positron and an electron, the entanglement is non-negligible. In addition, our calculation here does not consider the electron spin and its coupling effect with the angular momentum. A refined calculation should include it in the total Hamiltonian. The spin-angular momentum coupling is considered perturbation to the ground state derived from just the Coulomb interaction. One question is that whether the electron-nuclear entanglement in the ground state is still maintained when considering the spin-angular momentum perturbation. Further numerical calculation must be performed to answer this question. Finally, a more accurate treatment of this problem should employ the quantum field theory, as also pointed out in Ref.~\cite{Marcelo}. 

In this study, we are more interested in the case that the two subsystems are interacting through gravity potential. In this case, parameter $a$ is given by (\ref{GravityRadius}) and is called the Gravitational Bohr Radius. Its value is strongly depending on the values of masses $m_1$ and $m_2$. Considered the case of the Sun and the Earth. Given that $G=6.67\times 10^{-11} m^3kg^{-1}s^{-2}$, the mass of the Sun $m_1=1.99\times 10^{30}kg$, and the mass of the Earth $m_2=5.97\times 10^{24}kg$, and the Plank constant $\hbar = 1.06\times 10^{-34} m^2kgs^{-1}$, (\ref{GravityRadius}) gives $a=2.35\times 10^{-135}m$. This is much smaller than the Plank length $1.62\times 10^{-35}m$ and becomes non-physical. Clearly the wave function given in (\ref{GroundState}) is not suitable to describe quantum state of the Sun-Earth system. Consequently, it is meaningless to discuss entanglement between the quantum states of the Sun and the Earth.

However, suppose the masses of the two subsystems are at the scale of $~10^{-21}kg$ and $~10^{-17}kg$, such as the mass of a Brome mosaic virus~\cite{Maul97} and a vaccinia virus~\cite{Johnson}, respectively. One can estimate $a \simeq 5 cm$. If a universe consists only two such viruses and they interact only through classical gravity, and if we further assume such virus can be considered as quantum systems, then the conclusion can be drawn that the two systems are entangled when they are in the ground state. Obviously these are very strict conditions. Due to such practical limitation, one must be very cautious to draw such a conclusion. Instead, the significance of our result comes from the conceptual implication as discussed earlier.

\section{Entanglement Through Adjacent Interferometers}
\label{sec:Interferometers}
In an effort to confirm that gravitational field is a quantum entity, Refs~\cite{Bose, Marletto} proposed a novel experiment to induce entanglement between two test masses interaction through gravity. The motivation there is that entanglement can only be generated through mediation of a quantum entity. By confirming that mutual gravitational interaction between test masses can entangle the states of two masses, one can conclude that gravity field necessarily obeys quantum mechanics principles. The experiment is briefly restated below. Two test masses with masses $m_1$ and $m_2$ are prepared in superposition of two spatial separated states $|L\rangle$ and $|R\rangle$. Suppose the distance between the center of the two state is $\Delta x$. Each state is a localized Gaussian wave packets with the width much smaller than $\Delta x$, so that $\langle L|R\rangle = 0$. The distance between the centers of the two masses is $d$. Essentially these initial conditions can be physically realized by two Mach-Zehnder interferometers separated at distance $d$. The initial state is then given by
\begin{equation}
    \label{MZinit}
    |\Psi(t=0)\rangle_{12}=\frac{1}{\sqrt{2}}(|L\rangle_1+|R\rangle_1)\frac{1}{\sqrt{2}}(|L\rangle_2+|R\rangle_2).
\end{equation}
Now the two masses go through time evolution with mutual gravitational interaction for a period $\tau$. The time evolution introduces an additional phase shift in the probability amplitude, given by $\phi_{ij} = \frac{V(r_{ij})\tau}{\hbar}=\frac{Gm_1m_2\tau}{\hbar r_{ij}}$, where $i\in \{L, R\}$ is index for mass 1, $j\in \{L, R\}$ is index for mass 2, and $r_{ij}$ is the distance between the distinct components of the superposition state of the two masses. Since $r_{ij}$ is different for the four possible combinations of spatial states of the two masses, the final state is
\begin{equation}
    \label{MZfinal}
    \begin{split}
    |\Psi(t=\tau)\rangle_{12} & =\frac{e^{i\phi_{LL}}}{2}|L\rangle_1(|L\rangle_2+e^{i(\phi_{LR}-\phi_{LL})}|R\rangle_2) \\
    & +\frac{e^{i\phi_{RL}}}{2}|R\rangle_1(|L\rangle_2+e^{i(\phi_{RR}-\phi_{RL})}|R\rangle_2).
    \end{split}
\end{equation}
$\Psi(t=\tau)\rangle_{12}$ can be factorized if the following condition is met,
\begin{equation}
    \label{deltaphi}
    \Delta\phi = \phi_{LR}-\phi_{LL} + \phi_{RR}-\phi_{RL} = 2n\pi.
\end{equation}
For the two masses interact with classical gravity in the inteferometers, $r_{LL}=r_{RR}=d$, $r_{LR}=d+\Delta x$, and $r_{RL}=d-\Delta x$, we obtain
\begin{equation}
    \label{deltaphi2}
    \Delta\phi = \frac{Gm_1m_2\tau}{\hbar}(\frac{2}{d}-\frac{1}{d+\Delta x} - \frac{1}{d -\Delta x}),
\end{equation}
which is in general not equal to $2n\pi$ if proper parameters $d$ and $\Delta x$ are chosen. Thus, $|\Psi(t=\tau)\rangle_{12}$ cannot be factorized and entanglement between two test masses can be created. We now proceed to discuss the similarity and difference between this result and our result.

First, both results show that gravitational interaction can induce entanglement between two masses, and consequently confirm that gravity is a quantum entity if we acknowledge the reasoning described in Refs~\cite{Bose, Marletto}. However, the entanglement in the interferometer approach is generated through a very specific experimental design, the test masses are prepared in specific initial condition, while the result presented in this paper is generic and derived rigorously from first principle, i.e., from the Schr\"{o}dinger Equation. In this sense, our result generalizes the finding in Refs~\cite{Bose, Marletto} since it is not depending on specific experimental setup.

Second, both results show that the entanglement can be generated through other interaction such as Coulomb interaction. The key is that the interaction potential energy cannot be factorized into two independent terms with respective to the position degree of freedoms. To see this, suppose $V(r_{ij})=U(r_i)+W(r_j)$, it is easy to verify that $\Delta\phi =0$. One crucial example is the gravitational field from the Earth acting on the two masses, which can be approximated as $V(z_{ij}) = m_1gz_i + m_2gz_j$, where $z$ is the distance between the surface of the Earth and the masses. For this gravitational field, $\Delta\phi = 0$ and cannot induce entanglement. The same conclusion can be drawn from Eq.(\ref{SE}). When $V(r_{12})=U(r_1)+W(r_2)$, (\ref{SE}) can be separated two independent equations and admits $\psi(\vec{r}_1,\vec{r}_2)=\phi(\vec{r}_1)\varphi(\vec{r}_2)$ as a solution, which is a product state. Thus, the origin of the entanglement strongly depends on the geometry properties of the interacting field, although this is not pointed out in Refs~\cite{Bose, Marletto},

Third, one of the motivations of our work is to search for direct connection between entanglement and the geometry properties of spacetime, see detailed discussion in Section \ref{subsec:curvature}. This is a new insight not presented in Refs~\cite{Bose, Marletto}. The significance of the experiments proposed in Refs~\cite{Bose, Marletto} is that the gravity induced entanglement is practically detectable, while our work is mostly theoretical and conceptual.

\section{Discussions}
\label{sec:discussion}
\subsection{Entanglement and Spacetime Curvature}
\label{subsec:curvature}
The calculation result that two masses $A$ and $B$ interacting through gravitational field are entangled in the ground state has some interesting conceptual implications. First of all, we need to emphasize that the entanglement calculated in section IV is different from the entanglement in the AdS/CFT correspondence. The later refers to the correlation between a spatial region in the boundary of AdS and the rest of the boundary without gravity defined. On the other hand, the entanglement calculated in this paper is for two masses that interacts through classical gravitational field. 

Entanglement between $A$ and $B$ implies that there is correlation information between them. By knowing information on $A$, one can infer information on $B$. What information is correlated in the case of gravity interaction? The degree of freedom in our calculation is the position of the masses. Thus, the correlation encoded in the entangled systems is about the position of each system. Each system is in a mixed state. Their positions are correlated and cannot be described independently. Subsystem $A$ has to be described relative to subsystem $B$ for completeness~\footnote{More precisely, the bipartite system as a whole is in a pure state while subsystem $A$ is described relative to subsystem $B$.}. 

The root cause of such correlation is due to the intrinsic property of the gravitational potential, which is proportional to the inverse of relative distance between $A$ and $B$. The origin of such property can be better understood in the context of General Relativity. In GR, there is no gravitation field. Instead, the gravitation field in the Newtonian formulation is just the effect of spacetime curvature in the limit of very weak fields and low velocities~\cite{SeanCarroll}. Here, the full spacetime metrics, which determine the curvature, is written as $g_{\mu\nu} = \eta_{\mu\nu} + h_{\mu\nu}$, where $\eta_{\mu\nu}$ is the Minkowskian metrics and $h_{\mu\nu}$ is the small deviation on it. Only the $h_{00}$ component of the metrics is of important in the limit of very weak fields and low velocities, which turns out to be proportional to the inverse of relative distance between the two masses. Thus, the curvature of spacetime is manifested in such geometry property of the classical gravity potential. Since our calculation shows that the gravity with such property induced entanglement between two masses in their ground state, we argue that the curvature of spacetime causes the entanglement between the two masses. In other words, the entanglement at the ground state intrinsically connects to the curvature of spacetime. Given the first excited state is also an entangled state, we speculate that a quantum state for two masses interacting with gravity maybe always in an entangled state. Of course this statement is speculative only and needs a unified quantum gravity theory $\cal{M}$ for accurate treatment.

The argument that entanglement of two masses is connected to the curvature of spacetime the two masses create is not based first principles from a unified quantum gravity theory $\cal{M}$, since it is still being searched for. Instead, the argument is based on the generalized Bohr's Correspondence Principle~\cite{Bohr, Post}. Traditional quantum mechanics with classical gravity interaction, nevertheless, can be considered an approximation of theory $\cal{M}$. We expect when certain classical limit is imposed, theory $\cal{M}$ should either predict similar result as the less general physical theory, or explain why the results are different. In this notion, whether the ground state of two systems interacting through gravity is an entangled state can be a check point for theory $\cal{M}$ in the classical limit.

\subsection{Dependency on Reference Frame}
The Schr\"{o}dinger Equation in the form of (\ref{WF1}) implies that we have chosen a reference coordinate system such that the positions of the two systems are given by variable $\vec{r}_1$ and $\vec{r}_2$, respectively. The entanglement obtained in the calculation is with respect to such coordinate system. For instance, in the case of hydrogen atom, we can choose the lab where the hydrogen atom is prepared as a reference system. The coordinate system is at rest with respect to the lab. To study the entanglement properties in a relativistic framework, one should use the QFT. It has been shown that under Lorentz transformation, the entanglement measure for the subsystem-to-subsystem partition is Lorentz invariant~\cite{Calmet}\footnote{The reason for this is that the density matrix for a bipartite system is a Lorentz scalar. If the reduced density matrix is obtained by partitioning the density matrix with a Lorentz invariant term, the resulting reduced density matrix is again a Lorentz scalar. Thus, the entanglement measure is Lorentz invariant. See details in Ref.~\cite{Calmet}.}. In the case of a curve spacetime, whether the entanglement measure is invariant under coordinator transformation requires further analysis\footnote{For instance, under the covariant theory of quantum gravity~\cite{Cremaschini}, the Hamilton and wave equation can be formulated in the forms of 4-scalar, paving the path to calculate the entanglement measure. However, the calculation is beyond the scope of this paper.}. On the other hand, recent studies suggest that if we consider the reference system as quantum system as well, entanglement properties may depend on the choice a reference system~\cite{Brukner, Hoehn2018}. For example, we have shown that in a hydrogen atom, the electron and the nuclear are entangled in ground state with the lab as a reference frame. Consequently, the electron is in a mixed state. But if instead, we choose the nuclear itself as the reference system, the electron can be in a pure state with respect to the nuclear. Here the nuclear is considered as a quantum reference frame. These considerations are consistent with the calculation in this paper. Of course, the reason that entanglement properties depends on the choice a quantum reference system as shown in Refs.~\cite{Brukner, Hoehn2018} is more complex. Briefly speaking, by considering the reference frame as a quantum system, we need to describe the observed system and the reference system with relational properties. The entanglement properties in such a framework turn out to depend on the quantum reference system being chosen.  

\subsection{Ubiquitous Entanglement?}
Since the formulation and calculation in earlier sections are generic to any bipartite system where the interaction between two subsystems of is described as a central potential only depends on the distance between the two subsystems, the calculation is also applicable to well-studied systems such as a hydrogen atom where the interaction is described by the Coulomb potential, or a three-dimensional harmonic oscillator. In the case of a hydrogen atom, the entanglement in the ground state is an interesting fact because traditionally one is only interested in the energy levels of each eigenstate of a hydrogen atom. The entanglement information encoded in the wave function had been left unrecognized. Furthermore, given the ubiquity of gravitational interaction, one might think the entanglement between two masses are also ubiquitous. A quantum state for two masses interacting with gravity may not be separable. However, as pointed out in Section \ref{sec:limit}, for typical massive objects, the Gravitational Bohr Radius is much smaller than the Plank length. It is not meaningful to discuss entanglement between such massive objects. Furthermore, for a given mass object, when considering the gravitational interactions from virtually infinite other surrounding mass objects, the entangled state for a bipartite system could be disturbed such that the entanglement relation is destroyed. Thus, entanglement found in these well-studied bipartite systems can be practically ignored. The calculation and discussion in this paper is mostly for conceptual interest.

\section{Conclusions}
\label{sec:conclusion}
In this paper we study the entanglement measure of a bipartite system interacting through a central potential in the context of non-relativistic quantum mechanics. Thorough numerical calculation, we found that the ground state and the first excited state of such quantum system are entangled states. Although the entanglement properties of such systems were studied in earlier literature, this paper provides two contributions.

First, the formulation to calculate the entanglement measure in Section \ref{sec:entmeasusre} is more generic as contrary to earlier formulation~\cite{Tommasini} that is only applicable to state that are translationally invariant. 

Second, more importantly, we believe that the significance of the entanglement of such bipartite systems are not fully analyzed. Our motivation is to extract potential conceptual implications from this simple calculation, particularly in the case of gravitational interaction. For instance, the result provides hints of direct connection between entanglement and gravity. Since the gravitational field is an approximation of the curvature of spacetime in the limit of very weak fields and low velocities, it is reasonable to argue that there is an intrinsic connection between the entanglement of two masses and the curvature of the spacetime they create. A quantum gravity theory should either predict similar result in the weak field and low velocity limit, or gives a reasonable explanation why the results may be different.



%
%

\begin{acknowledgements}
The author thanks the anonymous reviewers for their careful reviews and valuable feedback that help to clarify the discussions and conclusions.
\end{acknowledgements}



\appendix
\section{Justification of Entanglement Measure}
Using the entanglement measure defined in (\ref{ent}), we say that two subsystems are entangled when $E>0$, which corresponds to $Tr(\hat{\rho}_1^2) < 1$. On the other hand, the two subsystems are separable when $E=0$, which corresponds to $Tr(\hat{\rho}_1^2) = 1$. We will show that $E$ indeed measures the entanglement of the bipartite system. Supposed the state vector of the bipartite system can be decomposed with a set of orthogonal basis $\{|\phi_i\rangle, |\varphi_i\rangle\}$ where $i=0, 1, 2,\ldots, d-1$, and $d$ could be infinite,
\begin{equation}
    \label{Schmidt}
    \begin{split}
    |\Psi\rangle_{12} & = \sum_i\lambda_i |\phi_i\rangle|\varphi_i\rangle \\ 
    & = \sum_i\lambda_i\{\int\phi_i(x)|x\rangle dx\}\{\int\varphi_i(y)|y\rangle dy\}\\
    & = \int\{\sum_i\lambda_i\{\phi_i(x)\varphi_i(y)\}|x\rangle|y\rangle dxdy,
    \end{split}
\end{equation}
where $\sum_i|\lambda_i|^2=1$. Compared to (\ref{WF1}) gives
\begin{equation}
    \label{Schmidt1}
    \psi(x,y) = \sum_i\lambda_i\phi_i(x)\varphi_i(y).
\end{equation}
The orthogonal relations are given by
\begin{equation}
    \int \phi_i(x)\phi^*_j(x)dx=\delta_{ij}, \int \varphi_k(y)\varphi^*_l(y)dy=\delta_{kl}.
\end{equation}
Substituting (\ref{Schmidt1}) into (\ref{purity2}) and applying the orthogonal properties, we obtain
\begin{equation}
    \label{purity10}
    Tr(\hat{\rho}_1^2) = \sum_{i=0}^{d-1}(|\lambda_i|^2)^2.
\end{equation}
Since $\sum_i|\lambda_i|^2=1$, it can be shown that
\begin{equation}
    \frac{1}{d^2} \leq Tr(\hat{\rho}_1^2) \leq 1.
\end{equation}
$Tr(\hat{\rho}_1^2)=1$ if and only if $d=1$, which implies $\psi(x,y)=\phi(x)\varphi(y)$ and consequently, $\Psi_{12}$ is a separable state. On the other hand, when $\psi(x,y)\ne\phi(x)\varphi(y)$, $d>1$ hence $Tr(\hat{\rho}_1^2) < 1$, we get $E>0$. Thus, (\ref{ent}) is a proper quantity to measure on whether the two subsystems are entangled. When $d\to\infty$, the lower bound of $Tr(\hat{\rho}_1^2)$ can be 0.

(\ref{Schmidt}) is essentially the Schmidt decomposition of the state vector for the bipartite system. However, it is not clear whether the decomposition is applicable when $d\to\infty$. Furthermore, it is very difficult to find the analytic solution of the decomposition in order to use (\ref{purity10}. Practically, we still rely on numerical method to calculate $ Tr(\hat{\rho}_1^2)$.

With (\ref{purity2}), $Tr(\hat{\rho}_1^2)$ is calculated in the $|x\rangle$ position basis, which strictly speaking is not a Hilbert space, because the norm is a delta function, i.e., $\langle x_i|x_j\rangle = \delta(x_i-x_j)$. We can calculate $Tr(\hat{\rho}_1^2)$ in the basis $\{|\phi_i\rangle\}$ instead, where the new basis form a truly Hilbert space since the norm is 1 by definition. Since $|\Psi\rangle_{12} = \sum_i\lambda_i |\phi_i\rangle|\varphi_i\rangle$, we can derive the reduced density operator
\begin{equation}
    \hat{\rho}_1 = Tr_2(|\Psi\rangle_{12}\langle\Psi|) = \sum_i|\lambda_i|^2|\phi_i\rangle\langle\phi_i|.
\end{equation}
From this one can derive the same expression of $Tr(\hat{\rho}_1^2)$ as (\ref{purity10}). In other words, quantity $Tr(\hat{\rho}_1^2)$, and consequently the entanglement measure $E$, are invariant in either the position basis or the basis $\{|\phi_i\rangle\}$. This is not surprised since the transformation between the two basis are unitary. The transform matrix element for variable $x$ can be written as
\begin{equation}
    \label{element}
    \begin{split}
    M_{ij} & = \langle x_i|\phi_j\rangle \\
    & = \{\int\delta(x-x_i)\langle x| dx\}\{\int\phi_j(x')|x'\rangle dx'\} \\
    &=\int\phi_j(x)\delta(x-x_i)dx = \phi_j(x_i).
    \end{split}
\end{equation}
Similarly, $M^{\dag}_{ki}=M^*_{ik}=\phi^*_k(x_i)$. Then,
\begin{equation}
    \label{element}
    \begin{split}
    (M^{\dag}M)_{kj}& =\sum_iM^{\dag}_{ki}M_{ij} = \int\phi^*_k(x_i)\phi_j(x_i)dx_i = \delta_{kj}.
    \end{split}
\end{equation}
This confirm $M^{\dag}M=I$ and $M$ is unitary.

\section{Harmonics Oscillator}
For a bipartite system that behaves like a harmonic oscillator, the potential energy is given by $V(\vec{r_1}-\vec{r_2}) = \frac{1}{2}\omega^2r^2_{12}$, where $\omega$ describes the strength of the potential. Using the center of mass coordinate system, the wave function for ground state is given by (\ref{transform2}) and
\begin{equation}
    \label{harmonicWF}
    \varphi_g(\vec{r}_{12}) = \sqrt{\frac{1}{\pi^{3/2} a^3}}e^{-\frac{1}{2}(\frac{r_{12}}{a})^2},
\end{equation}
where $a$ is constant determined by the masses and $\omega$.
Substituting this into (\ref{purity5}), using the same notations as for (\ref{purity7}), and denoting $\beta=L/(4a)=\alpha/4$, we obtain
\begin{equation}
\label{purity11}
\begin{split}
    Tr(\hat{\rho}_1^2) = & \lim_{\beta\to\infty}\frac{\beta^6}{\pi^3}\int_{-1}^{1} d\vec{\gamma'}_1d\vec{\gamma}_1 d\vec{\gamma'}_2d\vec{\gamma}_2 \\
    &\times e^{-2\beta^2(\gamma_{12}^2+\gamma_{1'2}^2+\gamma_{12'}^2+\gamma_{1'2'}^2)}.
\end{split}
\end{equation}
We can perform similar Monte Carlo calculation on this twelve-dimensional integration. But fortunately, this integration can be calculated analytically, so that the result can be used to check the accuracy of Monte Carlo integration. First, we expand $\gamma_{12}^2 = (\bar{x}_1-\bar{x}_2)^2+(\bar{y}_1-\bar{y}_2)^2+(\bar{z}_1-\bar{z}_2)^2$, where we denote dimensionless variables $\bar{x}_1=2x_1/L$, $\bar{x}_2=2x_2/L$ and so on. Also noted that $d\vec{\gamma'}_1=d\bar{x}_1d\bar{y}_1d\bar{z}_1$, (\ref{purity11}) can be simplified into
\begin{equation}
\label{purity12}
\begin{split}
    Tr(\hat{\rho}_1^2) = & \lim_{\beta\to\infty}\{\frac{\beta^2}{\pi}\int_{-1}^{1} d\bar{x}'_1d\bar{x}_1 d\bar{x}'_2d\bar{x}_2 \\
    &\times e^{-2\beta^2((\bar{x}_1-\bar{x}_2)^2+(\bar{x}'_1-\bar{x}_2)^2+(\bar{x}_1-\bar{x}'_2)^2+(\bar{x}'_1-\bar{x}'_2)^2)}\}^3\\
    & = \lim_{\beta\to\infty}\{\frac{\beta^2}{\pi}\int_{-1}^{1}J^2(\bar{x}_2,\bar{x}'_2)d\bar{x}'_2d\bar{x}_2\}^3, 
\end{split}
\end{equation}
where
\begin{equation}
\label{Jfunction}
    J(\bar{x}_2,\bar{x}'_2) = \int_{-1}^{1} e^{-2\beta^2((\bar{x}_1-\bar{x}_2)^2+(\bar{x}_1-\bar{x}'_2)^2)}d\bar{x}_1.
\end{equation}
Expanding the exponent in the integral for the $J$ function, $(\bar{x}_1-\bar{x}_2)^2+(\bar{x}_1-\bar{x}'_2)^2 = 2(\bar{x}_1 - \breve{x})^2 + (\bar{x}_2-\bar{x}'_2)^2/2$ where $\breve{x}=(\bar{x}_1+\bar{x}_2)/2$,
\begin{equation}
\label{Jfunction2}
\begin{split}
     J(\bar{x}_2,\bar{x}'_2) & = e^{-\beta^2(\bar{x}_2-\bar{x}'_2)^2} \int_{-1}^{1}e^{-4\beta^2(\bar{x}_1 - \breve{x})^2}d\bar{x}_1 \\
     &=\frac{\sqrt{\pi}}{4\beta}e^{-\beta^2(\bar{x}_2-\bar{x}'_2)^2}\times \\
     & [Erf(2\beta(1-\breve{x})) - Erf(e\beta(1+\breve{x}))].
\end{split}
\end{equation}
Here $Erf(x)=\pi^{-1/2}\int^x_{-x}e^{-t^2}dt$ is  the error function. Since $-1< Erf(x) < 1$, the difference of the error function at two arbitrary values $a$ and $b$ is, $-2< [Erf(a) - Erf(b)]< 2$. Thus, $J^2(\bar{x}_2,\bar{x}'_2) < \frac{\pi}{4\beta^2}e^{-2\beta^2(\bar{x}_2-\bar{x}'_2)^2}$. Plug this into (\ref{purity12}),
\begin{equation}
    \label{purity13}
    Tr(\hat{\rho}_1^2) < \lim_{\beta\to\infty}\{\frac{1}{4}\int_{-1}^{1}e^{-2\beta^2(\bar{x}_2-\bar{x}'_2)^2}d\bar{x}'_2d\bar{x}_2\}^3.
\end{equation}
Recall $\beta = L/(4a)$, $\alpha = L/a$, and denote $\grave{x}_2 = \alpha\bar{x}_2/2$, $\grave{x}'_2 = \alpha\bar{x}'_2/2$, we rewrite (\ref{purity13}) as
\begin{equation}
    Tr(\hat{\rho}_1^2) < \lim_{\alpha\to\infty} K^3(\alpha),
\end{equation}
where
\begin{equation}
    \label{K}
     K (\alpha) = \frac{1}{\alpha^2}\int_{-\alpha/2}^{\alpha/2}e^{-\frac{1}{2}(\grave{x}_2-\grave{x}'_2)^2}d\grave{x}'_2d\grave{x}_2.
\end{equation}
Denote $t=((\grave{x}_2-\grave{x}'_2))/\sqrt{2}$, $s=(\alpha/2-\grave{x}_2)/\sqrt{2}$, and using the error function again, 
\begin{equation}
    \label{K2}
    \begin{split}
        K (\alpha)&=\frac{\sqrt{2}}{\alpha^2}\int_{-\alpha/2}^{\alpha/2}\{\int_{-(\alpha/2-\grave{x}_2)/\sqrt{2}}^{(\alpha/2-\grave{x}_2)/\sqrt{2}}e^{-t^2}dt\}d\grave{x}_2\\
        & = \frac{2\sqrt{\pi}}{\alpha^2}\int_{0}^{\alpha/\sqrt{2}}Erf(s)ds
    \end{split}
\end{equation}
Note that $Erf(s)$ is a monotonically increasing function and approaches 1 rapidly when $s$ is a large enough number, e.g., $Erf(2)=0.9953$, $Erf(3)=0.9999$, we can approximate the integral $\int_{0}^{\alpha/\sqrt{2}}Erf(s)ds \simeq \alpha/\sqrt{2} - \delta$ where $\delta$ is some positive finite number. Thus, 
\begin{equation}
    \label{K3}
    K (\alpha) \simeq \frac{2\sqrt{\pi}}{\alpha^2}(\frac{\alpha}{\sqrt{2}} - \delta).
\end{equation}
We ignore $\delta$ when $\alpha\to\infty$, and finally get the upper bound of the purity
\begin{equation}
    \label{purity14}
    Tr(\hat{\rho}_1^2) < \lim_{\alpha\to\infty}\frac{(2\pi)^{3/2}}{\alpha^3} \to 0.
\end{equation}
However, $Tr(\hat{\rho}_1^2) \geq 0$, we conclude that $Tr(\hat{\rho}_1^2)=0$. 

Table \ref{tab:4} shows the Monte Carlo estimation of the twelve-dimensional integration in (\ref{purity11}). The last column of the table lists the value of $K^3(\alpha)$. These results confirm that the Monte Carlo integration is consistent with the analytic results. 
\begin{table}[h!]
\caption{Purity of $\rho_1$ at Ground State}
\label{tab:4}       
\renewcommand{\arraystretch}{1.2}
\begin{tabular}{m{1cm}|m{1.8cm}|m{1.8cm}|m{1.3cm}|m{1.8cm}}
\hline
$L/a$ & $Tr(\hat{\rho}_1^2)$ & Error & $N_{MC}$ (million) & $(\sqrt{2\pi}/\alpha)^3$\\
\hline
10 & $1.25\times10^{-2}$ & $4.99\times 10^{-4}$ & 8 & $1.57\times 10^{-2}$\\
20 & $1.06\times 10^{-3}$ & $6.83\times 10^{-5}$ & 128 & $1.97\times 10^{-3}$\\
40 & $1.50\times 10^{-5}$ & $2.75\times 10^{-6}$ & 256 & $2.46\times 10^{-4}$\\
60 & $1.19\times 10^{-6}$ & $1.41\times 10^{-7}$ & 512 & $7.29\times 10^{-5}$\\
100 & $1.31\times 10^{-7}$ & $2.45\times 10^{-8}$ & 1,024 & $1.57\times 10^{-5}$\\
\hline
\end{tabular}
\renewcommand{\arraystretch}{1}
\end{table}
\end{document}